# Peer interaction facilitates co-construction of knowledge in quantum mechanics

Mary Jane Brundage, Alysa Malespina, and Chandralekha Singh

*Department of Physics and Astronomy, University of Pittsburgh, Pittsburgh, PA 15260*

**Abstract.** Collaborative learning with peers can lead to students learning from each other and solving physics problems correctly not only in situations in which one student knows how to solve the problems but also when none of the students can solve the problems alone. We define the rate of construction as the percentage of groups collaborating on problem-solving that solve the problem correctly out of all groups having at least one member who answered correctly and one incorrectly while solving the same problem individually first. We define the rate of co-construction on each problem as the percentage of collaborating groups that answered it correctly if no student in the group individually answered it correctly before the collaborative work. In this study, we investigated student learning measured by student performance on a validated quantum mechanics survey and rates of construction and co-construction of knowledge when students first worked individually after lecture-based instruction in relevant concepts and then worked with peers during class without receiving any feedback from the course instructor. We find that construction of knowledge consistently occurred at a high rate during peer collaboration. However, rates of co-construction were more varied. High rates of co-construction were generally achieved when approximately half of the students knew the correct answers initially. We also conducted an analysis of some of the survey questions that correlate with high rates of co-construction to gain some insight into what students converged on after peer interaction and what types of difficulties were reduced. Our findings can be valuable for instructors who want to provide in-class and out-of-class opportunities for peer collaboration in their physics courses.

## I. INTRODUCTION AND THEORETICAL FRAMEWORK

Classroom time is often one of the most constraining factors for any instructor looking to improve their instruction and student outcomes. In quantum mechanics (QM) courses, instructors may not have as many students as they do in introductory courses, but students must learn an entire new paradigm of physics in a limited time. Thus, in addition to other support provided by the instructor, incentivizing and providing opportunities for students to learn from their peers both inside and outside of the classrooms may be an effective tool for helping them develop a good grasp of QM concepts.

Independent student collaboration without involvement of instructors has been recommended by influential educators and psychologists for over a century. The influential psychologist Dewey's framework of participatory democracy encourages a learning environment in which there is collective ("associated") participation of all students [1] (p. 47), cooperation between peers [1] (p. 72), and opportunities to develop intellectually in new (i.e., outside of their home and family) social environments [1] (p. 44). All these goals can be facilitated through independent group work in which students are allowed to construct knowledge collaboratively, even without instructor intervention [1] (pp. 48-49). Additionally, keeping in mind the limited capacity of human working memory, the work of Hutchens describes distributed cognition, in which learners use outside resources to stretch their cognitive resources and accomplish intellectual tasks [2, 3]. This framework includes the use of tools (such as a calculator or written list of formulas) as well as social distribution of cognition when students collaborate with each other to circumvent the limited capacity of an individual's working memory while solving problems [2-4]. This social distribution increases cognitive capacity not only by allowing students to combine their knowledge base (increasing the capacity of working memory), but by allowing them to build on each other's ideas, organize work and communicate ideas [4].

Contemporary researchers have also found benefits in peer collaboration and proposed more nuanced frameworks. One framework created specifically for pairs of students, the Zone of Proximal Facilitation (ZPF) model [5], predicts that collaborative success can be achieved when participants have some knowledge or skill in the domain but the task at hand is just beyond their ability to complete alone. When students are faced with such a challenge, they must participate in deliberate problem solving and apply their current knowledge to a new task in which that learning can be transferred [5, 6]. Although student expertise related to a domain is generally on a spectrum, this framework distinguishes novice and expert students in a particular domain as those at the two ends of this expertise spectrum. Within this model, novice students are those who can sometimes complete simple tasks (for example, fact retrieval or straightforward concept application) on their own but cannot independently complete complex tasks (for example, transfer of knowledge from one situation to another). When collaborating, two novice students are likely to succeed at basic tasks and may sometimes also be able to complete complex tasks. Expert students can independently complete simple tasks and sometimes complete complex tasks but they also benefit from working with other students (novice or expert) on complex problems. However, according to ZPF model, students at a certain level of expertise may not benefit from collaboration if the task is beyond the scope of their group competence, but when two students collaborate on tasks within their group competence, they are likely to be able to solve complex problems [5, 6].

Inspired by these historic and contemporary theoretical frameworks advocating peer collaboration even in the absence of instructor support and feedback during the collaborative work, there is a critical need for research on how group work can improve student learning in physics courses as measured by students' improved performance in different situations. Students in physics courses may benefit from peer collaboration for a variety of reasons. For example, if they all share the prior knowledge useful for solving a problem, students may benefit from peer interaction and collaborative work if some of the group members cue relevant prior knowledge that all students did not individually identify as relevant to the problem. Students are also more likely to remember what they discussed and learned together than alone [5, 7, 8]. Alternatively, if none of the students initially identified all of the relevant information or approaches useful to solve a problem, they may have adequate combined knowledge to be able to do it correctly consistent with the framework of distributed cognition and Zone of Proximal Facilitation model [9]. Not only are students likely to benefit from checking the logic and rationale of each other's solutions while collaborating with their peers [10], they are also likely to think more deeply about the logic of their own solution when explaining their approach to others. These deeper reflections may reveal inaccuracies in their initial reasoning and may help students co-construct knowledge. Thus, if classroom instructional time is limited, out-of-class peer collaboration can enhance and complement classroom instruction, freeing instructors to spend their time on other issues including assisting students with new or challenging content and skills for which students need instructors' support [11].

The benefits of peer collaboration are not limited to combining student knowledge bases and extending the capacity of each individual's working memory. Students are also often more comfortable explaining their own reasoning and questioning others' reasoning when speaking to a peer rather than an authority figure, so they are likely to gain more practice with scientific communication and critical thinking when working collaboratively with peers [12-19]. In particular, an added benefit of problem solving and learning via peer collaboration is that students are less likely to be anxious when communicating their thoughts with peers, as anxiety can take up cognitive resources that students could otherwise dedicate to the task at hand [20-22]. Additionally, peer collaboration can also positively affect students' motivational beliefs such as self-efficacy [23-25], which has been shown to positively correlate with student performance and persistence in STEM courses [26-29].

Peer collaboration has been shown to be effective in the context of college physics instruction [30-34]. Mazur's peer instruction method has been widely used successfully to help students develop a solid grasp of physics concepts, and involves short but frequent instances of collaboration during class with the help of multiple-choice questions [30, 35]. This formative assessment approach has been shown to be effective in improving student learning. It can help keep students engaged in the learning process during the lecture because students know that they will be held accountable by peers, and requires that students think about how they will explain to peers their own views and thought processes regarding physics [30, 35]. Prior research shows that self-efficacy can play a role when students discuss concepts with each other and explain concepts to others to cement their knowledge of a concept [32].

Physics instructors have also used peer collaboration to help students learn effective problem solving strategies while solving complex, real-world problems [36, 37] and to help students learn physics in an active learning style without lectures [38-40]. When students solve real-world problems [36, 37], they are likely to think explicitly about the applicability of physics concepts in those contexts and reflect upon their problem solving process, which are conducive to learning effective problem solving strategies while solidifying their knowledge structure.

Previous work from our group has also shown that peer collaboration can result in better performance on introductory physics surveys, such as the Conceptual Survey of Electricity and Magnetism (CSEM), than if the students worked on the CSEM individually [41, 42]. Moreover, after the group work, when students were administered the CSEM individually, their average performance working individually was comparable to the average group performance, which implies that students retained what they learned with peers. Moreover, we found that the group work benefitted students regardless of whether they worked in a group or individually first. We also found that regardless of what their initial performance was, students on average, significantly benefited from group work compared to those who did not work with a peer. We also found that in unaided peer collaboration after lecture-based instruction on relevant concepts, students were able to co-construct knowledge approximately 30% of the time when working with a peer. In particular, in cases in which no student correctly answered a question correctly, as a group, they were able to come to the correct answer 30% of the time [41, 42]. Students are likely to be able to co-construct knowledge while collaborating with peers for several reasons. For example, if they initially chose different incorrect answers, they are likely to explain their reasoning to each other to determine who is correct. This explanation may reveal flaws in their initial approach and lead to co-construction of knowledge. Alternatively, students who selected the same incorrect answer may admit to their peers that they have doubts about their approach. This can lead to productive discussions among peers about the correct approaches to problem solving and lead to co-construction of knowledge.

Inspired by the frameworks supporting effectiveness of peer collaboration even without any intervention from the instructor, we studied the impact of peer collaboration on construction and co-construction of knowledge as measured by student performance in the first semester of their quantum mechanics physics courses. We used the Quantum Mechanics Formalism and Postulates Survey (QMFPS) [43], a validated survey, to investigate the effectiveness of peer interaction for upper-level undergraduate and first-year graduate physics students after traditional lecture-based instruction in relevant concepts. Although prior research on collaborative work involving three or more students [36, 37, 44, 45], has found it beneficial to assign roles to students collaborating with each other (e.g., group leader, scribe, skeptic, time-keeper etc.), prior research also suggests that there are significant benefits to letting students select their partners if students are asked to collaborate in pairs [46, 47]. Since our investigation involved mainly pairs of students (except in two cases when students were in a group of three), we let students select who they wanted to collaborate with on brainstorming how to answer the QMFPS questions correctly.

It is not clear a priori how much students will benefit from unguided peer discussions during their collaboration particularly because the paradigm of QM differs from introductory physics and has the potential to affect peer interaction. For example, students in introductory mechanics courses often have common alternative conceptions since the laws of physics are not consistent with everyday intuitive ideas, e.g., a huge trailer truck would exert a larger force on a small car or an object moving at a constant velocity must have a net force acting on it. Thus, since students who have not mastered introductory mechanics are likely to have similar incorrect ideas, e.g., about force and motion, discussions with peers may be effective for helping them reconcile the differences between those alternative conceptions and ideas consistent with Newton's laws of motion. On the other hand, one may assume that students taking QM are unlikely to have common alternative conceptions because the quantum paradigm is very different from what students experience in everyday experiences. However, even though quantum mechanics is abstract and nonintuitive, students often have common difficulties in learning this new paradigm and research shows that they sometimes transfer their ideas from classical to quantum contexts or from one quantum context to another quantum context in which those ideas are not applicable [48-50]. In other words, even in the quantum mechanics contexts, students' reasonings in different contexts show patterns, which can impact the manner in which students are likely to construct and co-construct knowledge during unguided peer collaboration. We also note that the group of students who take quantum mechanics may have a stronger sense of community than introductory students, as they are upper-level undergraduate or graduate students who have either shared experiences in prior classes or are taking multiple classes together during the semester they collaborated on answering the QMFPS survey questions discussed here. This connection is likely to positively impact the outcomes of collaboration.

This study investigates the extent to which construction and co-construction of knowledge occurs when students collaborate with their peers in groups of two (sometimes three) in the new context of QM courses using a validated survey that focuses on the formalism and postulates of quantum mechanics after having worked on it individually after traditional instruction in relevant concepts. We analyze rates of construction and co-construction of knowledge. Here, we define construction as the instance in which only one student answered a question correctly individually before working with a peer, and then as a group they were able to converge on the correct answer. We also investigate specific features of some survey items that may have led to high rates of construction or co-construction on those items. Specifically, we aim to answer the following research questions regarding the QMFPS when physics upper-level undergraduate and graduate students in quantum mechanics courses worked in pairs (sometimes three students together) after working individually after instruction in relevant concepts:

RQ1. Does working in a group improve student performance on the QMFPS?
RQ2. How often do students in groups choose a correct answer if one student (but not both) chose the correct answer individually, which we define as construction of knowledge?
RQ3. How often do students in groups choose a correct answer if no student in that group chose the correct answer individually, which we define as co-construction of knowledge?
RQ4. What item features correlate with high rates of construction and co-construction of knowledge for some of the survey items?

## II. Methods

### A. Participants and Procedures

All participants attended a large public university, which is a large, public, and urban research institution in the US. This research was carried out in accordance with the principles outlined in the university's Institutional Review Board (IRB) ethical policy. Almost all students were either enrolled in the first mandatory course in a two-semester undergraduate quantum mechanics course for physics juniors/seniors or in a mandatory first-year first semester graduate introduction to teaching course in the physics department. Almost all students had received instruction on all topics covered by the survey in their undergraduate courses before taking the QMFPS survey. For the undergraduate course, the QMFPS survey was administered at the end of the first-semester course which is mandatory for all physics majors and covers chapters 1-4 of Griffiths' Introduction to Quantum Mechanics [51], first individually and then in groups in the following class period. In the first-year first semester graduate teaching of physics course that helps graduate teaching assistants reflect upon evidence-based approaches to teaching and learning, the survey was administered in the first few weeks of the semester to help graduate students (most of whom were also teaching assistants for various introductory physics courses) reflect upon the value of collaborative learning. Although these graduate students were simultaneously enrolled in the first core graduate quantum mechanics course for Ph.D. students in physics, since the QMFPS survey was administered in the first few weeks of the semester for the graduate students, their main learning of the QMFPS content was in their undergraduate courses similar to the undergraduate students who participated in the study. Student test performance, construction, and co-construction from the graduate and undergraduate groups were very similar, so we provide statistics on combined data.

Students (N=78) first took the QMFPS individually and were given a full 50-minute class period to complete the survey. Then, in the next scheduled class, students took the survey a second time in groups of two (N=36) or three (N=2) without any assistance from the instructor. Students were given a full 50-minute class period to collaborate on the survey. Students were told that they will take a quiz in both classes, but not that the second one was the same as the first and that they will be working with a peer. In particular, they were not told that they will be working with a peer for the second quiz until the next scheduled class in which they took the QMFPS survey again with a peer. This was to control for the "positive effect resulting from increased attention and recognition of the subjects" when they work together with their peers on the same survey in the next scheduled class [52]. Moreover, they were not given feedback on their performance between the first and second survey attempts, so they did not know if their initial responses were correct.

### B. Survey

The Quantum Mechanics Formalism and Postulates Survey (QMFPS) is a validated survey which includes topics that were not focused on in other quantum mechanics surveys [43], such as Dirac notation, Hilbert space, state vectors, physical observables and their corresponding Hermitian operators, compatible and incompatible observables, projection operators and writing operators in terms of their eigenstates and eigenvalues, and spin angular momentum. The final version of the survey contains 34 multiple-choice items and is designed primarily for junior or senior-level undergraduates and first-year graduate students. It has been validated [43] using initial faculty and student interviews, learning from previous open-ended and multiple-choice questions given to QM students at the university, and additional faculty and student interviews for each iteration of the survey. Interviews served multiple purposes, including assuring that students interpreted questions as intended and development of alternative incorrect choices for multiple-choice questions aligned with common student difficulties when the questions were presented as open-ended [43].

### C. Analysis

First, for each question, we calculated the percent of students who selected the correct answer individually, as well as the percent of student groups that answered the question correctly. We also analyzed the rates of "construction" and co-construction for each question. Here we define the rate of construction as the percentage of groups that choose the correct answer out of all groups having at least one member who answered correctly and one incorrectly. When looking at the combined individual and group scores, we write the binary scores (where 0 is incorrect and 1 is correct) of each student and then the group as a whole. For example, assuming groups with two members, a group that constructed knowledge can be written as 011 or 101. In special cases involving groups of three, one can eliminate the repeated answer choice for individual scores. The rate, or percentage, of construction on each question can be found as follows:

$$\frac{N(101) + N(011)}{N(100) + N(010) + N(101) + N(011)} x100\%$$

(1)

We define the rate of co-construction on each question as percentage of groups that choose the correct answer if no students in the group individually chose the correct answer. Assuming groups of two, group that co-constructed knowledge can be written as 001. The rate, or percentage, of co-construction on each question is:

$$\frac{N(001)}{N(000) + N(001)} x100\%$$

(2)

A more detailed analysis of construction and co-construction which also accounts for the groups of three students appears in Appendix A.

Analyzing questions individually allows us to find patterns in the data such as correlations between the percentage of individuals that answered the questions correctly individually and the rate of co-construction. Next, we analyzed the average rate of construction and co-construction for questions in content-based groups, such as quantum states, commutation relations and Dirac notation. This allowed us to find any content-based patterns, for example, do students have particularly high rates of co-construction for questions regarding commutation relations? Finally, we analyze the content of some of the questions with particularly high rates of construction or co-construction individually. This allows us to find potential qualitative patterns in questions that may facilitate productive peer interaction.

### III. Results and Discussion

*RQ1. Does working in a group improve student performance on the QMFPS?*

Overall, the average score for individual students on the QMFPS was 53% (standard deviation (SD) = 18%), and the average score for groups was 72% (SD = 15%). We define normalized gain, $g$, to compare individual (Pre) and group (Post) scores, as $g = (Post-Pre)/(100- Pre)$.

**Table I.** The percentages of individuals and groups that chose the correct answer, normalized gain, and rate of construction and co-construction for each question. If there is no number in the co-construction column, there were no groups in which both students initially answered incorrectly. Individual and group scores are rounded in the table and normalized gains were calculated before rounding.

| Item # | Individual | Group | Normalized Gain | Construction | Co-Construction |
|---|---|---|---|---|---|
| 1 | 50 | 68 | 37 | 92 | 8 |
| 2 | 69 | 84 | 49 | 93 | 0 |
| 3 | 45 | 71 | 47 | 100 | 8 |
| 4 | 87 | 95 | 59 | 80 | - |
| 5 | 54 | 68 | 32 | 70 | 33 |
| 6 | 37 | 53 | 25 | 67 | 20 |
| 7 | 56 | 74 | 40 | 76 | 25 |
| 8 | 33 | 66 | 49 | 95 | 25 |
| 9 | 44 | 58 | 25 | 64 | 21 |
| 10 | 81 | 87 | 32 | 86 | 0 |
| 11 | 73 | 92 | 71 | 100 | 0 |
| 12 | 69 | 84 | 49 | 81 | 25 |
| 13 | 67 | 87 | 61 | 88 | 40 |
| 14 | 64 | 79 | 41 | 80 | 0 |
| 15 | 71 | 87 | 55 | 84 | 0 |
| 16 | 37 | 55 | 29 | 76 | 19 |
| 17 | 60 | 82 | 54 | 80 | 67 |
| 18 | 53 | 76 | 50 | 71 | 55 |
| 19 | 54 | 74 | 43 | 80 | 30 |
| 20 | 47 | 74 | 50 | 77 | 44 |
| 21 | 79 | 87 | 36 | 70 | 33 |
| 22 | 26 | 37 | 15 | 75 | 13 |
| 23 | 88 | 84 | - | 57 | - |
| 24 | 64 | 76 | 34 | 80 | 17 |
| 25 | 44 | 61 | 30 | 76 | 9 |
| 26 | 69 | 92 | 74 | 90 | 50 |
| 27 | 56 | 87 | 70 | 85 | 71 |
| 28 | 49 | 55 | 13 | 63 | 17 |
| 29 | 18 | 47 | 36 | 85 | 28 |
| 30 | 27 | 39 | 17 | 69 | 10 |
| 31 | 48 | 76 | 53 | 93 | 25 |
| 32 | 35 | 66 | 47 | 88 | 27 |
| 33 | 45 | 76 | 56 | 93 | 44 |
| 34 | 30 | 59 | 41 | 73 | 38 |

As seen in Table I, one item had negative gain (so we do not calculate the normalized gain for it as is customary), seven items had low gain ($g < 30\%$, [53]), twenty-four items had medium gain $30\% \leq g < 70\%$, [53]), and three items had high gain $g \geq 70\%$, [53]). Thus, most questions (79%) showed either medium or large normalized student learning gains. The only item that had negative gain was item 23, which had a very high individual average of 88% and a group average of 84%. No other item had an individual average score of over 90% (see Table I), so peer collaboration may not be useful if almost all students have already mastered the material.

We note that while one may assume that students performance will improve when taking the QMFPS survey a second time in group after working individually, it is important to note that students did not receive any feedback between their individual and group attempts. The improved group performance supports the claim that working in groups is useful to students over and above any practice effect due to working on the QMFPS survey twice. Past work in our group [41, 42] found that when students in an introductory electricity and magnetism course were divided into two randomly assigned groups and half of the students first worked individually and then collaborated with a peer and the other half first collaborated with a peer and then worked individually, on average, there were no differences between group performances in the two cases. Moreover, in the same study, when group work was followed by individual work, student individual performance was on average comparable to the group performance. In other words, on average, students retained what they learned from working with a peer.

Also, low or high rate of improvement from individual to group QMFPS survey may be due to a variety of reasons. For example, a low rate of improvement may be due to either high scores on the individual survey so that there is little room for improvement, or poor understanding of the content so that students could not figure out how to solve the problems correctly even with the help from peers. Thus, while it is important to note that students improved when working in a group with peers, improvement in scores may not be as informative. Instead, rates of construction or co-construction may be better measures for determining the concepts for which group work was particularly effective.

*RQ2. How often do students in groups choose a correct answer if some students chose the correct answer individually?*

While the exact equation that includes groups of 2 and 3 is given earlier, if all groups consisted of 2 students then if one student in a group answered a question correctly on the individual survey, and the other student in that group answered the question incorrectly, but the group answered the question correctly, we call this situation peer construction of knowledge ("construction"). The rates of construction for each question can be seen in Table I. Here, we analyze trends in some questions and topics that had the highest rates of construction. The average rate of construction for each individual question was varied and ranged from 57% (Question 23) to 100% (Questions 3 and 11). However, the overall average rate of construction is quite high (80%, SD=11%), and the correlation between construction and individual survey scores was non-significant ($r = 0.02$, $p = 0.90$) [54]. Consistently high rates of construction among survey questions add to the existing evidence that unguided group work is a valuable tool for student learning in a variety of circumstances [23, 38, 39, 55]. Additionally, the rates of construction were remarkably similar between topics (for example "measurement" or "Dirac notation", see Table II). All topics had between 74% and 85% construction, as seen in Table II.

**Table II.** One possible concept-based categorization of the survey questions, the number of questions that fall in each category, and the item numbers belonging to each category. The table also includes the combined rate of construction and co-construction for each topic. When considering multiple-questions related to a given topic, we use Eqs. (1) and (2) where N is for all questions related to a topic. The number

of questions in different categories does not add up to 34 because some questions fall into more than one category.

| Topic | Item number(s) | Construction | Co-construction |
| --- | --- | --- | --- |
| Quantum states | 1, 4, 7, 11-15, 17, 19 | 83 | 22 |
| Eigenstates of operators corresponding to physical observables | 1, 4, 7, 14, 15, 17, 18, 20 | 79 | 31 |
| Time development of quantum states | 3-7, 26, 32, 34 | 80 | 25 |
| Measurement | 2-5, 7-9, 13, 19, 21, 23-25, 27, 28, 31-34 | 81 | 26 |
| Expectation value of observables | 5, 10, 22 | 74 | 15 |
| Time dependence of expectation value of observables | 15-17, 29, 30 | 79 | 21 |
| Commutators/compatibility | 16, 17, 19, 20, 27, 28 | 77 | 33 |
| Spin angular momentum | 20-30 | 77 | 23 |
| Dirac notation | 4, 9-14, 18 | 81 | 27 |
| Dimensionality of the Hilbert space | 1 | 85 | 8 |

It is also important to address the cases in which one student in a group answered a question correctly on the individual survey, the other student answered the question incorrectly, and the group answered the question incorrectly. This may happen for a variety of reasons. For example, a student with a dominant personality might convince others that their incorrect logic is correct. We did find that in some cases, if both students chose the correct answer individually, the group answered the question incorrectly. In these cases, group work had a negative effect on student performance. For twenty-six of the thirty-four items, there was no instance of this happening. For seven items, this occurred in one group, and for one item this happened in 3 of the groups. Because it was rare for peer interaction and collaboration to result in decreased student performance, this unguided peer collaboration activity has very small risk for students and even though the group work was unguided by an instructor, the benefits of construction outweigh the risks of one student convincing the other of an incorrect answer being correct.

*RQ3. How often do students in groups choose a correct answer if no students chose the correct answer individually?*

While the exact equation that includes groups of 2 and 3 is given earlier, if all groups consisted of 2 students then if neither student in a group answered a question correctly on the individual survey, and the group answers the question correctly, we call this peer co-construction of knowledge ("co-construction").

The average rate of co-construction for each individual question can be seen in Table I, and was quite varied, ranging from 0% (Questions 2, 10, 11, 14, and 15) to 71% (Question 27). Table I reveals that the highest rates of co-construction were for questions that approximately 50-60% of students answered correctly on the individual survey. This is consistent with Mazur's established Peer Instruction method guidelines, which suggest that most effective peer instruction occurs when approximately half of students can answer a question correctly individually [35]. Similarly, the lowest rates of co-construction tend to be for clusters of items that had individual scores furthest from the 50-60% range. Mazur's observation aligns with the Zone of Proximal Development Framework. For questions that few students were able to answer correctly (for example, Question 29, which Table I shows 18% of students answered correctly individually) they may not know the necessary concepts to answer the question correctly in a group, even if their cognitive resources and skills are increased by working with a partner [5, 6]. Alternatively, for questions that most students were able to answer individually (for example, Question 23, which Table I shows 88% of students answered correctly), since such a high percentage of students may understand the material, there is little room for improvement via peer collaboration [5, 6].

Rates of construction and co-construction for topics, rather than individual questions, can be found in Table II. The rates of co-construction were highest for commutators and compatibility (33%) and eigenstates of operators corresponding to physical observables (31%). The rates of co-construction were lowest for dimensionality of Hilbert space (8%, though there was only one question on this topic), and expectation values of observables (15%).

*RQ4. What item features correlate to high rates of construction and co-construction of knowledge on some of the survey items?*

Below, we focus on some of the survey items with high rates of co-construction in order to highlight a wide range of quantum mechanics topics on which students benefited from peer interaction. Although question 18 had a high rate of co-construction, it is left out of this analysis because slightly different versions of the answer choices were given to different classes. The questions we discuss below with a high rate of co-construction are: Questions 13, 17, 20, 26, 27, 33, and 34. By analyzing these questions, we can learn about the aspects of these questions that led to their high co-construction rates and improved group performance. The percentages of students and groups that chose each answer option for these featured survey questions can be found in Table III. Table IV in Appendix B provides performance for all questions.

Analysis of items 13, 17, 20, 26, 27, 33, and 34 shows several features that may lead to high rates of co-construction for these items. First, most of these high co-construction questions had a mix of incorrect answers selected. In particular, questions 13, 17, 20, and 33 did not have a "dominant" incorrect answer choice but had a mix of incorrect answer choices that each had under 20% of the students selecting them. We hypothesize that this mix of incorrect answers may have been a result of a certain level of speculation that led to fruitful student discussions and learning when working in groups.

**Table III.** For each item of interest, the percentage of students or groups that chose each answer option. The correct answer is bold, underlined, and italicized. Some questions have fewer responses than others because some instructors did not give questions 31-34 of the survey. Students who skipped a question were marked incorrect on that question.

| Item # | Group/Individual | N | A | B | C | D | E | Construction | Co-Construction |
|---|---|---|---|---|---|---|---|---|---|
| 13 | Individual | 78 | 3 | 5 | 14 | 12 | **67** | 88 | 40 |
|    | Group      | 38 | 0 | 5 | 3  | 5  | **87** |    |    |
| 17 | Individual | 78 | **60** | 8 | 3 | 14 | 15 | 80 | 67 |
|    | Group      | 38 | **82** | 5 | 0 | 5  | 8  |    |    |
| 20 | Individual | 78 | 22 | **47** | 4 | 4 | 22 | 77 | 44 |
|    | Group      | 38 | 8  | **74** | 0 | 3 | 16 |    |    |
| 26 | Individual | 78 | 5 | 15 | 3 | **69** | 8 | 90 | 50 |
|    | Group      | 38 | 3 | 5  | 0 | **92** | 0 |    |    |
| 27 | Individual | 78 | 38 | 1 | **56** | 0 | 3 | 85 | 71 |
|    | Group      | 38 | 13 | 0 | **87** | 0 | 0 |    |    |
| 33 | Individual | 60 | 12 | 12 | **45** | 7 | 25 | 93 | 44 |
|    | Group      | 29 | 7  | 0  | **76** | 0 | 17 |    |    |
| 34 | Individual | 60 | 53 | 12 | **30** | 5 | 0 | 73 | 38 |
|    | Group      | 29 | 38 | 3  | **59** | 0 | 0 |    |    |

Item 13 focuses on student knowledge of quantum states, measurement, and Dirac notation, and had the following text and answer choices (the correct answer is in bold):

Question 13. Choose all of the following statements that are correct.

(1) $|\Psi\rangle = \int \langle p|\Psi\rangle |p\rangle dp$
(2) $|\Psi\rangle = \int \Psi(x)|x\rangle dx$
(3) If you measure the position of the particle in the state $|\Psi\rangle$, the probability of finding the particle between $x$ and $x + dx$ is $|\langle x|\Psi\rangle|^2 dx$.

A. 1 only    B. 1 and 2 only.    C. 1 and 3 only.    D. 2 and 3 only.    **E. all of the above.**

Table III shows that the most common individual answer was the correct one, and the most common incorrect answer choices were options C and D. Thus, most students understood that statement 3 was correct, but some only recognized that either 1 or 2 was correct, but not both. Prior work [56] suggests that a common student difficulty is converting between the Dirac notation and wavefunction representation, (e.g., in position representation $|\Psi\rangle = \Psi(x) = \langle x|\Psi\rangle$, and the analogous relation in the momentum representation). Table III also shows that students struggled equally to select options C and D as correct. Because Dirac notation is new for most taking QM courses, the high rate of co-construction (40%) may at least partly be evidence for the power of distribution of cognition when learning new concepts. By combining working memory and relevant knowledge together, students may be able to figure out all the relevant knowledge needed to convert between different representations and answer the question correctly.

Item 17 investigates student understanding of conserved quantities in quantum mechanics and focuses on commutators/compatibility, eigenstates of operators corresponding to physical observables and whether they commute with the Hamiltonian of the system. It has the following text and answer choices (the correct answer is in bold):

Question 17. Choose all of the following statements that are <u>necessarily</u> correct.
(1) An observable whose corresponding time-independent operator commutes with the time-independent Hamiltonian of the system, $\hat{H}$, corresponds to a conserved quantity (constant of motion).
(2) If an observable $Q$ does not depend explicitly on time, $Q$ is a conserved quantity.
(3) If a quantum system is in an eigenstate of the momentum operator at initial time $t = 0$, momentum is a conserved quantity.

**A. 1 only**  B. 2 only  C. 3 only  D. 1 and 3 only  E. all of the above

Though Table III shows that the most common answer individually was the correct one, i.e., option A, options D and E were the most common incorrect individual choices. This means that most students knew that statement 1 is true. Previous interviews [57] found that many students incorrectly claimed that if the system is in an eigenstate of an operator, the corresponding observable is a conserved quantity. This belief may make statement 3 attractive to students. Moreover, for free particles, for which the Hamiltonian commutes with the momentum operator, momentum is a conserved quantity. Prior research also shows that many students do not always differentiate between the eigenstates of the Hamiltonian operator and those of other operators (e.g., momentum operator) [57]. With regard to the choice E, i.e., "all of the above", students who selected this option overgeneralized that all time-independent observables correspond to conserved quantities. Question 17 has a high rate of co-construction (67%). We hypothesize that one reason peer collaboration may be particularly valuable in this context is that students may benefit from collaboration if they understand that observables whose corresponding operators commute with the Hamiltonian of the system are conserved, but have difficulty differentiating between or applying ideas related to conserved quantities in other situations.

Item 20 investigates student knowledge of eigenstates and eigenvalues of operators corresponding to physical observables in the context of spin angular momentum as well as concepts related to simultaneous eigenstates of two operators. It had the following text and answer choices (the correct answer is in bold):

Question 20. For a spin-1/2 particle, suppose $|s, m_s\rangle = \left|\frac{1}{2}, -\frac{1}{2}\right\rangle$ is a simultaneous eigenstate of $\hat{S}^2$ and $\hat{S}_z$ with quantum numbers $s = \frac{1}{2}$, and $m_s = -\frac{1}{2}$. Choose all of the following statements that are correct.
(1) $\hat{S}_+ \left|\frac{1}{2}, -\frac{1}{2}\right\rangle$ is an eigenstate of both $\hat{S}^2$ and $\hat{S}_z$.
(2) If $\hat{S}^2 \left|\frac{1}{2}, -\frac{1}{2}\right\rangle = \frac{3}{4}\hbar^2 \left|\frac{1}{2}, -\frac{1}{2}\right\rangle$, then $\hat{S}_+ \left|\frac{1}{2}, -\frac{1}{2}\right\rangle$ is an eigenstate of $\hat{S}^2$ with eigenvalue $\frac{3}{4}\hbar^2$.
(3) If $\hat{S}_z \left|\frac{1}{2}, -\frac{1}{2}\right\rangle = -\frac{\hbar}{2}\left|\frac{1}{2}, -\frac{1}{2}\right\rangle$, then $\hat{S}_+ \left|\frac{1}{2}, -\frac{1}{2}\right\rangle$ is an eigenstate of $\hat{S}_z$ with eigenvalue $-\frac{\hbar}{2}$.

A. 1 only  **B. 1 and 2 only**  C. 1 and 3 only  D. 2 and 3 only  E. all of the above

Table III shows that the most common individual answer was option B, the correct answer. However, options A and E were also common choices. Thus, most students knew statement 1 to be true, though some students did not realize that statement 2 was true or that statement 3 was false when answering individually. These difficulties were reduced after peer collaboration, and this item had a co-construction rate of 44%. Our prior interviews with some students during the validation of the QMFPS survey suggest that some students struggled with raising and lowering operators and did not realize that $\hat{S}_+ \left|\frac{1}{2}, -\frac{1}{2}\right\rangle$ yields an eigenstate of $\hat{S}^2$ with eigenvalue $\frac{3}{4}\hbar^2$ (statement 2) but it yields an eigenstate of $\hat{S}_z$ with eigenvalue $\hbar/2$. We also note that $\hat{S}_z \left|\frac{1}{2}, -\frac{1}{2}\right\rangle = -\frac{\hbar}{2}\left|\frac{1}{2}, -\frac{1}{2}\right\rangle$ was provided to students earlier in the survey in the instructions but some students may have overlooked the information provided.

Item 26 investigates student knowledge of time development of quantum states in the context of spin angular momentum, and had the following text and answer choices (the correct answer is in bold):

Question 26. The Hamiltonian of a charged particle with spin-1/2 at rest in an external uniform magnetic field is $\hat{H} = -\gamma B_0 \hat{S}_z$ where the uniform field $B_0$ is along the z-direction and $\gamma$ is the gyromagnetic ratio (a constant). Suppose that at time $t = 0$, the particle is in an initial normalized spin state $|\chi\rangle = a\left|\frac{1}{2}, \frac{1}{2}\right\rangle + b\left|\frac{1}{2}, -\frac{1}{2}\right\rangle$ where $a$ and $b$ are suitable constants. What is the state of the system after time $t$?

A. $|\chi(t)\rangle = e^{\frac{i\gamma B_0 t}{2}} \left(a\left|\frac{1}{2}, \frac{1}{2}\right\rangle + b\left|\frac{1}{2}, -\frac{1}{2}\right\rangle\right)$

B. $|\chi(t)\rangle = e^{\frac{-i\gamma B_0 t}{2}} \left(a\left|\frac{1}{2}, \frac{1}{2}\right\rangle + b\left|\frac{1}{2}, -\frac{1}{2}\right\rangle\right)$

C. $|\chi(t)\rangle = e^{\frac{i\gamma B_0 t}{2}} \left((a+b)\left|\frac{1}{2}, \frac{1}{2}\right\rangle + (a-b)\left|\frac{1}{2}, -\frac{1}{2}\right\rangle\right)$

**D. $|\chi(t)\rangle = ae^{\frac{i\gamma B_0 t}{2}} \left|\frac{1}{2}, \frac{1}{2}\right\rangle + be^{\frac{-i\gamma B_0 t}{2}} \left|\frac{1}{2}, -\frac{1}{2}\right\rangle$**

E. None of the above.

Individually, most students chose the correct answer (option D, 69%), which can be seen in Table III. The rate of co-construction was high (50% of eligible cases). The most common incorrect answer was option A, but very few groups chose this answer after peer collaboration. The only difference between answers options A and D is that option A does not acknowledge the different time-dependent phase factors between the spin up and spin down components of the state vector. Previous research on similar items [50] found that this difficulty often is due to students struggling to distinguish between time-evolution of energy eigenstates (or stationary states) and their linear combinations and thinking that the time evolution of a quantum state is always via an overall time-dependent phase factor.

Item 27 investigates student knowledge of measurement and commutators/compatibility, and had the following text and answer choices (the correct answer is in bold):

Question 27. Suppose that at time $t = 0$, the particle is in an initial state in which the $x$-component of spin $S_x$ has a definite value $\frac{\hbar}{2}$. Choose all of the following statements that are correct about measurements performed on the system starting with this initial state at $t = 0$.
(1) If you measure $S_x$ immediately following another measurement of $S_x$ at $t = 0$, both measurements of $S_x$ will yield the same value $\frac{\hbar}{2}$.
(2) If you first measure $\vec{S}^2$ at $t = 0$ and then measure $S_x$ in immediate succession, the measurement

of $S_x$ will yield the value $\frac{\hbar}{2}$ with 100% probability.
(3) If you first measure $S_z$ at $t = 0$ and then measure $S_x$ in immediate succession, the measurement of $S_x$ will yield the value $\frac{\hbar}{2}$ with 100% probability.

A. 1 only    B. 3 only    **C. 1 and 2 only**    D. 1 and 3 only    E. 2 and 3 only

Table III shows that the most common answer individually was option C, which is correct, but option A was a popular incorrect answer (38% of students individually chose this). Most students understood that statement 1 is true, but many students did not realize that statement 2 is also true. Our prior interviews with some students suggest that they had difficulty identifying statement 2 as true because they either did not realize that $\vec{S}^2$ and $S_x$ commute, or they did not realize that commuting variables can share a complete set of simultaneous eigenstates. After working in groups, 13% of students still did not realize this. However, this question had a high rate of co-construction (71%).

Item 33 investigates student knowledge of measurement of different observables (energy and position), and had the following text and answer choices (the correct answer is in bold):

Question 33. The wavefunction at time $t = 0$ is $\Psi(x,0) = \frac{\psi_1(x)+\psi_2(x)}{\sqrt{2}}$. Choose all of the following statements that are correct <u>at time $t = 0$</u>:
(1) If you measure the position of the particle at time $\underline{t = 0}$, the probability density for measuring $x$ is $\left|\frac{\psi_1(x)+\psi_2(x)}{\sqrt{2}}\right|^2$.
(2) If you measure the energy of the system at time $\underline{t = 0}$, the probability of obtaining $E_1$ is $\left|\int_0^a \psi_1^*(x)\left(\frac{\psi_1(x)+\psi_2(x)}{\sqrt{2}}\right)dx\right|^2$.
(3) If you measure the position of the particle at time $\underline{t = 0}$, the probability of obtaining a value between $x$ and $x + dx$ is $\int_x^{x+dx} x|\Psi(x,0)|^2\, dx$.

A. 1 only    B. 3 only    **C. 1 and 2 only**    D. 1 and 3 only    E. All of the above

The most popular individual response, which can be seen in Table III, was the correct answer (C), but 25% of student chose option E, and 12% of students each chose options A and B. After working collaboratively with peers, most students converged on the correct answer, though a minority of groups still chose option E. The most common confusion was thinking that statement 3 is correct, followed by thinking statement 2 is incorrect. Prior work related to this question [56] suggests that some students who thought that statement 3 was correct confused the probability of measuring position of the particle between x and x + dx with the expectation value of position (though the integral in statement 3 does not have the correct limits for the expectation value). Prior research also suggests that students often have difficulties distinguishing between probability and expectation value [50, 56, 58]. What is also interesting is that the statements 1 and 3 asked for the probability density and the probability of measuring position which should only differ by "dx," but the given expression in statement 3 refers to a common alternate conception (there is an extra factor of x) and many students did not explicitly check consistency of their responses for their selections of both statements 1 and 3 as correct (which is not possible). Moreover, prior research suggests that when asked about the probability of measuring energy for this state, some students recognize that the probabilities of measuring the ground state and first excited state energies are ½ each but they have difficulty recognizing that the expression in statement 2 would yield ½ since it is the square of the scalar product of the given quantum state of the system with the ground state written in position representation. We note that this expression in statement 2 is more difficult for students to recognize in position representation than in Dirac notation, if students are familiar with the Dirac notation. Prior interviews also suggest that some students may have not realized that statement 2 is correct because they thought that the expression for the probability of measuring energy $E_1$ should include the Hamiltonian operator [56].

Item 34 also investigates student knowledge of the measurement of different observables (energy and position), and had the following text and answer choices (the correct answer is in bold):

Question 34. The wavefunction at time $t = 0$ is $\Psi(x, 0) = \frac{\psi_1(x) + \psi_2(x)}{\sqrt{2}}$. Choose all of the following statements that are correct <u>at a time $t > 0$</u>:
(1) If you measure the position of the particle <u>after a time $t$</u>, the probability density for measuring $x$ is $\left|\frac{\psi_1(x) + \psi_2(x)}{\sqrt{2}}\right|^2$.
(2) If you measure the energy of the system <u>after a time $t$</u>, the probability of obtaining $E_1$ is $\left|\int_0^a \psi_1^*(x)\left(\frac{\psi_1(x) + \psi_2(x)}{\sqrt{2}}\right) dx\right|^2$.
(3) If you measure the position of the particle <u>after a time $t$</u>, the probability of obtaining a value between $x$ and $x + dx$ is $\int_x^{x+dx} x|\Psi(x, 0)|^2\, dx$.

A. None of the above     B. 1 only     **C. 2 only**    D. 3 only    E. 1 and 3 only

Table III shows that individually, most students chose the incorrect answer, option A. Only 30% of students chose the correct answer, option C, individually. However, most groups correctly answered this question, and the co-construction rate was 38%. Most students chose option A individually, so the most common incorrect conception was that statement 2 is false. This statement is correct using the same logic as for Question 33, but students must also understand that the probabilities of measuring different values of energy do not change with time. This is because the Hamiltonian is a constant of motion and time-dependent phase factors will always cancel out in the probabilities of measuring different values of energy (or any observable whose corresponding operator commutes with the time-independent Hamiltonian of the system so that they are constants of motion).

Next, we divide these high co-construction items discussed into two categories. The first may be categorized as consisting of questions in the Zone of Proximal Facilitation (ZPF) model [5, 6], because they require straightforward application of concepts, such as questions 13, 20, and 26. These questions require knowledge of foundational concepts such as Dirac notation, eigenstates and eigenvalues of spin operators that commute with each other, and time evolution of quantum states, respectively. With regard to taxonomies of cognitive achievement such as Bloom's or Marzano's taxonomies, these questions are lower than the other questions discussed here [59, 60]. Using the ZPF model, these rates of co-construction may be high because students may either be able to cue each other about relevant knowledge, or they may combine their incomplete knowledge bases to construct and converge on the correct answer [5, 6].

Another category consists of questions that require complex, multistep solutions which are higher on Bloom's or Marzano's taxonomies. In the ZPF model, these would be labeled complex problems. These include items 17, 27, 33, and 34. In these questions, students must combine their base knowledge of formalisms and postulates in ways that require multiple steps or combine multiple concepts simultaneously. For example, in item 17, students needed to know what conserved quantities are and other related issues (why observables whose corresponding operators commute with the Hamiltonian of the system are special with regard to issues related to time-dependence) and in items 33 and 34, students needed to know how to find the probability of measuring energy in the position representation and the probability of measuring position for a given quantum state at time $t = 0$ and at later times. We hypothesize that in complex questions like these, the high rate of co-construction may at least partly be due to students taking advantage of distribution of cognition between peers, which may have allowed students to process more information during problem solving using their peers' cognitive resources as an external aid to converge on the correct reasoning [2-6].

## IV.    Summary and Conclusion

In this study, we found that advanced physics students in quantum mechanics courses who initially worked individually on the QMFPS survey improved their performance when working in groups even without any instructor feedback between individual and group attempts. We also found that construction (i.e., a group choosing a correct answer when only one or some of the students in the group chose the correct answer individually) occurred frequently on the survey. Because construction rates were high across all questions, and the risk of decreased learning outcomes from group work overall were low, our findings suggest that unguided peer collaboration in which some students know the correct solutions is a useful tool for instructors to use in their courses both inside and outside of the classroom.

We found that co-construction (i.e., a group choosing a correct answer when no students in the group chose the correct answer individually) occurred for most questions, but the rate varied more than for construction. The highest rates of co-construction were found for questions that approximately 50-60% of students answered correctly on the individual administration. Thus, co-construction may work best for concepts that are not extremely difficult or easy for most students. In the former case, two students may not have the combined knowledge to solve the problem and in the latter, there may not be significant opportunities for co-construction.

Finally, we conducted an analysis of some of the individual items with the highest rates of co-construction. We found co-construction may work particularly well for items that focus on fundamentals of quantum mechanics that may be challenging for students. These include Dirac notation, eigenstates and eigenvalues of spin operators that mutually commute with each other, compatible observables, and time evolution of quantum states. These concepts are new for many students, and we found that students benefited from peer interaction. Also, co-construction may work well for items that require many steps or many related concepts to answer correctly, e.g., those related to expressions for the probability of measuring energy in position representation or probability of measuring position. In these instances, also, we found that students benefited from peer collaboration, which may at least partly be due to distributing the cognitive load with peers. Considering students can co-construct knowledge in the realm of quantum mechanics, as shown in this study, and also in introductory physics in the context of electricity and magnetism [41], we believe that co-construction of knowledge is likely to occur in other introductory and advanced physics courses as well.

Importantly, peer collaboration provided students an opportunity to articulate their thoughts and understand their peer's thought processes. These types of opportunities can help students develop ability to communicate physics concepts in a low anxiety environment. Unfacilitated peer collaboration does not require extensive effort from instructors and carries minimal risk for students. Thus, this is an effective tool that instructors can implement even outside of their classes by providing students appropriate incentives, e.g., grade incentives, even though in this investigation, students worked with peers in the classroom.

**Acknowledgments**

We thank the National Science Foundation for award PHY-1806691. We thank students who participated in this study and Prof. R. P. Devaty for helpful suggestions and corrections to the manuscript.

## Appendix A: General Formula for Construction and Co-construction

Here we present, with mathematical rigor, the calculation of construction and co-construction rates.

First, for each question, we calculated the percent of students who selected the correct answer individually, as well as the percent of student groups that answered the question correctly. We also analyzed the rates of "construction" and "co-construction" for each question defined as follows.
We first define the notation $N(i,j;k)$ where
$i$ = number of students who answered the question correctly while working individually
$j$ = number of students in the group
$k$ = indicator of work as a group: *1* for groups which answered correctly (working as a group), *0* for groups which answered incorrectly.
$N(i,j;k)$ = the number of groups having $j$ members, $i$ of which answered the question correctly working individually, who answered the question working as a group correctly or incorrectly, based on the value of index $k$.

Using this notation, for each question, the rate of construction $R_{con}$ is defined as the fraction of groups having at least one member who answered correctly and one incorrectly when working individually who answered correctly while working as a group:

$$R_{con} = \sum_{j \geq 2} \sum_{i=1}^{j-1} N(i,j;1) \Big/ \sum_{k=0}^{1} \sum_{j \geq 2} \sum_{i=1}^{j-1} N(i,j;k)$$

(A1)

For our case, since the number of students in a group was 2 (most groups) or 3 (only for two groups), an expansion of this equation can be written by looking at the total number of groups that fit each performance criteria:

$$R_{con} = \frac{N(1,2;1) + N(1,3;1) + N(2,3;1)}{N(1,2;0) + N(1,3;0) + N(2,3;0) + N(1,2;1) + N(1,3;1) + N(2,3;1)}$$

(A2)

Using this notation, for each question, the rate of co-construction $R_{co-con}$ is defined as the fraction of groups having no members who answered correctly while working as individuals who answered correctly while working as a group:

$$R_{co-con} = \sum_{j \geq 2} N(0,j;1) \Big/ \sum_{k=0}^{1} \sum_{j \geq 2} N(0,j;k)$$

(A3)

When the groups have either 2 or 3 members, the result is:

$$R_{co-con} = \frac{N(0,2;1) + N(0,3;1)}{N(0,2;0) + N(0,3,0) + N(0,2;1) + N(0,3;1)}$$

(A4)

Analyzing questions individually allows us to find patterns in the data such as correlations between the percentage of individuals that answered the questions correctly individually and the rate of co-construction. Next, we analyzed the average rate of construction and co-construction for questions in content-based groups, e.g., quantum states and Dirac notation. This allowed us to find any content-based patterns, e.g., do students have particularly high rates of co-construction for questions regarding commutation relations? Finally, we analyze the content of some of the questions with particularly high rates of construction or co-construction individually. This allows us to find potential qualitative patterns in questions that may facilitate productive peer interaction.

**Appendix B: Individual and Group Performances as well as Construction and Co-construction Rates**

Table IV. For each item of interest, the percentage of students or groups that chose each answer option. The correct answer is printed is bold, underlined, and italicized. Some questions have fewer responses than others because some instructors did not give questions 31-34 of the survey. Students who skipped a question were marked incorrect on that question. For question 18, * denotes that the distribution of answer choices reflects the responses of 33 individual students and 16 groups who were administered the final version of QMFPS (but the construction and co-construction rates are for all students).

| Item # | Group/Individual | N | A | B | C | D | E | Construction | Co-Construction |
|---|---|---|---|---|---|---|---|---|---|
| 1 | Individual | 78 | **50** | 4 | 10 | 3 | 33 | 92 | 8 |
|   | Group | 38 | **68** | 5 | 3 | 5 | 18 |  |  |
| 2 | Individual | 78 | **69** | 1 | 9 | 17 | 4 | 93 | 0 |
|   | Group | 38 | **84** | 0 | 5 | 11 | 0 |  |  |
| 3 | Individual | 78 | 4 | 15 | 1 | 35 | **45** | 100 | 8 |
|   | Group | 38 | 0 | 3 | 0 | 26 | **71** |  |  |
| 4 | Individual | 78 | **87** | 4 | 5 | 3 | 1 | 80 | - |
|   | Group | 38 | **95** | 3 | 0 | 0 | 3 |  |  |
| 5 | Individual | 78 | 6 | 3 | 6 | **54** | 31 | 70 | 33 |
|   | Group | 38 | 3 | 0 | 3 | **68** | 26 |  |  |
| 6 | Individual | 78 | 4 | 4 | 21 | **37** | 35 | 67 | 20 |
|   | Group | 38 | 0 | 0 | 13 | **53** | 34 |  |  |
| 7 | Individual | 78 | 15 | 5 | 0 | **56** | 23 | 76 | 25 |
|   | Group | 38 | 16 | 3 | 3 | **74** | 5 |  |  |
| 8 | Individual | 78 | 37 | 17 | **33** | 5 | 6 | 95 | 25 |
|   | Group | 38 | 21 | 8 | **66** | 3 | 3 |  |  |
| 9 | Individual | 78 | 3 | 41 | 9 | 4 | **44** | 64 | 21 |
|   | Group | 38 | 3 | 32 | 5 | 3 | **58** |  |  |
| 10 | Individual | 78 | 5 | 0 | 3 | **81** | 10 | 86 | 0 |
|    | Group | 38 | 3 | 3 | 0 | **87** | 8 |  |  |
| 11 | Individual | 78 | 21 | 0 | **73** | 3 | 4 | 100 | 0 |
|    | Group | 38 | 8 | 0 | **92** | 0 | 0 |  |  |
| 12 | Individual | 78 | 4 | 24 | **69** | 0 | 3 | 81 | 25 |
|    | Group | 38 | 0 | 13 | **84** | 0 | 3 |  |  |
| 13 | Individual | 78 | 3 | 5 | 14 | 12 | **67** | 88 | 40 |
|    | Group      | 38  | 0  | 5  | 3  | 5  | **87** |    |    |
| 14 | Individual | 78  | **64** | 1  | 24 | 10 | 0  | 80 | 0  |
|    | Group      | 38  | **79** | 0  | 11 | 8  | 3  |    |    |
| 15 | Individual | 78  | 4  | **71** | 12 | 3  | 10 | 84 | 0  |
|    | Group      | 38  | 0  | **87** | 3  | 0  | 11 |    |    |
| 16 | Individual | 78  | 15 | 6  | 26 | 15 | **37** | 76 | 19 |
|    | Group      | 38  | 5  | 0  | 26 | 13 | **55** |    |    |
| 17 | Individual | 78  | **60** | 8  | 3  | 14 | 15 | 80 | 67 |
|    | Group      | 38  | **82** | 5  | 0  | 5  | 8  |    |    |
| 18 | Individual | 78* | 58 | 9  | 3  | 6  | **24** | 71 | 55 |
|    | Group      | 38* | 44 | 0  | 0  | 6  | **50** |    |    |
| 19 | Individual | 78  | 10 | 10 | 24 | 1  | **54** | 80 | 30 |
|    | Group      | 38  | 3  | 8  | 13 | 3  | **74** |    |    |
| 20 | Individual | 78  | 22 | **47** | 4  | 4  | 22 | 77 | 44 |
|    | Group      | 38  | 8  | **74** | 0  | 3  | 16 |    |    |
| 21 | Individual | 78  | 0  | 4  | 1  | **79** | 14 | 70 | 33 |
|    | Group      | 38  | 0  | 3  | 0  | **87** | 11 |    |    |
| 22 | Individual | 78  | 12 | 12 | 9  | 40 | **26** | 75 | 13 |
|    | Group      | 38  | 8  | 8  | 8  | 39 | **37** |    |    |
| 23 | Individual | 78  | 8  | **88** | 4  | 0  | 0  | 57 | -  |
|    | Group      | 38  | 13 | **84** | 3  | 0  | 0  |    |    |
| 24 | Individual | 78  | **64** | 1  | 12 | 19 | 4  | 80 | 17 |
|    | Group      | 38  | **76** | 11 | 5  | 8  | 0  |    |    |
| 25 | Individual | 78  | 18 | 8  | 19 | **44** | 12 | 76 | 9  |
|    | Group      | 38  | 18 | 3  | 8  | **61** | 11 |    |    |
| 26 | Individual | 78  | 5  | 15 | 3  | **69** | 8  | 90 | 50 |
|    | Group      | 38  | 3  | 5  | 0  | **92** | 0  |    |    |
| 27 | Individual | 78  | 38 | 1  | **56** | 0  | 3  | 85 | 71 |
|    | Group      | 38  | 13 | 0  | **87** | 0  | 0  |    |    |
| 28 | Individual | 78  | 4  | **49** | 13 | 31 | 3  | 63 | 17 |
|    | Group      | 38  | 3  | **55** | 11 | 26 | 3  |    |    |
| 29 | Individual | 78  | 6  | 3  | **18** | 46 | 26 | 85 | 28 |
|    | Group      | 38  | 3  | 0  | **47** | 32 | 16 |    |    |
| 30 | Individual | 78  | **27** | 1  | 6  | 55 | 8  | 69 | 10 |
|    | Group      | 38  | **39** | 0  | 3  | 53 | 3  |    |    |
| 31 | Individual | 60  | 7  | 8  | **48** | 13 | 23 | 93 | 25 |
|    | Group      | 29  | 7  | 3  | **76** | 0  | 14 |    |    |
| 32 | Individual | 60  | 38 | 13 | **35** | 7  | 7  | 88 | 27 |
|    | Group      | 29  | 28 | 3  | **66** | 0  | 3  |    |    |

| 33 | Individual | 60 | 12 | 12 | **45** | 7 | 25 | 93 | 44 |
|    | Group      | 29 | 7  | 0  | **76** | 0 | 17 |    |    |
| 34 | Individual | 60 | 53 | 12 | **30** | 5 | 0  | 73 | 38 |
|    | Group      | 29 | 38 | 3  | **59** | 0 | 0  |    |    |